\documentclass[aps,prl,twocolumn,preprintnumbers]{revtex4}
\usepackage{graphicx,amsmath}
\usepackage{dcolumn}
\usepackage{amsmath}
\usepackage{amssymb}
\usepackage{amsfonts}

\begin{document}

\title{A Note on the Equivalence of Gibbs Free Energy and Information Theoretic Capacity}         % Enter your title between curly braces
\author{David Ford, Physics NPS}        % Enter your name between curly braces
\email[]{dkford@nps.edu}
\affiliation{Department of Physics, Naval Post Graduate School, Monterey, California}
\date{\today}          % Enter your date or \today between curly braces
\begin{abstract}
The minimization of Gibbs free energy is based on the changes in work and free energy that occur in a physical or
chemical system. The maximization of mutual information, the capacity, of a noisy channel is determined based on
the marginal probabilities and conditional entropies associated with a communications system. 
As different as the procedures might first appear, through the exploration of a simple, ``dual use'' Ising model,
it is seen that the two concepts are in fact the same. In particular, the case of a binary symmetric channel is 
calculated in detail.
\end{abstract}
\maketitle

\section{Introduction}
In 1876 J.W. Gibbs  \cite{gibbs} proposed a refinement to the notion of Helmholtz free energy. The utility of the refinement  rests on the observation that
when attempting to determine the maximum amount of work that may be extracted from a thermodynamic process, some of that work may already
be accounted for, for example, work against the atmosphere $-p \Delta V$. The suggestion was that perhaps the actual quantity of interest
is the maximum amount of work {\it other than atmospheric} that may be extracted. This general concept, equipped with a {\it system dependent} form for the work term \cite{callen},\cite{l&l},\cite{krevet},\cite{kim}
is well known and widely used in physics, physical chemistry and many engineering disciplines.

Some seventy years later C. Shannon and W. Weaver \cite{s&w} provide a sound theoretical framework for determining  the maximum
amount of information that may be transmitted from sender to receiver through a noisy channel. Their concept, the channel capacity,
calculated by extremization of differences of various communication system entropies, is reminiscent of statistical physics \cite{jaynes}
but apparently different\cite{feynman}.

The purpose of this letter is to show how these two concepts are in fact  the same. 
To accomplish this it is useful to have a simple example that belongs to both the physics and communications theory 
traditions. The Ising model for magnetic systems is a good candidate. The literature on the topic is vast. Physical applications of the model
relevant to the present purpose include \cite{b&k}, \cite{chin&hong},\cite{f&h}. Relevant applications of the model in communications include \cite{nishimori},\cite{n&w},\cite{tanaka}. After a brief review of the Ising model, it will be shown that the real space renormalization procedure \cite{kadanoff} connects the Gibbs free energy and the 
channel capacity in a natural way. 

\section{Review}
As noted in the introduction, the Ising model is of interest in both physics and communications.
The states described by the model are a sequence of  electron spin up ($\uparrow$) or spin down ($\downarrow$) states
which are easily translated to the 1's and 0's of a bit stream. The simplest communications scheme is one in which the transmitted bits
are statistically independent of each other but coupled to the bits received.
\begin{figure}[htbp]
\begin{center}
\leavevmode
\includegraphics[width=50mm,keepaspectratio]{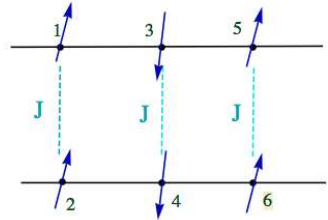}
\caption{A simple send/receive Ising model. The sent bits, $\sigma_1$, $\sigma_3$, $\sigma_3$ (odd numbered) are independent of
each other but coupled to their received bits  $\sigma_2$, $\sigma_4$, $\sigma_6$ (even numbered).}
\label{model1}
\end{center}
\end{figure}
\noindent The sequence shown in figure \ref {model1} consists of six
sites representing $2^6$ possible {\it configurations} of spins (bits). 

Following the paradigm of Boltzmann and Gibbs, the probability weight of each spin configuration is a function of its own particular energy as well as the energy of its environment, represented by the temperature \cite{hill}, \cite{huang}. Although at first the pedagogy may seem specialized to a narrow class of systems occurring in physics, engineering and chemistry, the approach is quite general and any discrete probability distribution $\{p_1, p_2, \ldots, p_N \}$ can be put into Boltzmann-Gibbs form \cite{ford}. 

The assumed statistical independence of the send/receive pairs manifests itself in terms of the physics
through the form of the energy function. 
Under these conditions, the energy of a configuration, $H_{\sigma1,\ldots,\sigma6}$, is comprised of additive contributions 
from the distinct send/receive pairs with no interaction energy terms
$$
H_{\sigma1,\ldots,\sigma6}= H_{\sigma_1 \, \sigma_2}+H_{\sigma_3 \, \sigma_4}+H_{\sigma_5 \, \sigma_6} .
$$
For this reason, to understand the properties of the total system, it is enough to study the properties of a
single send/receive pair. A typical Ising {\it Hamiltonian}, appropriate for a single send/receive unit, for example site 1 (send) and site 2 (receive) as shown in figure \ref{model1}, is of the form

$$ H_{\sigma_1 \sigma_2}=-J \sigma_{1} \sigma_{2} +   h ( \sigma_{1} + \sigma_{2} ) + c$$
\noindent with spin coupling $J$ and external field $h$.
The so-called Zeeman energy term, $c$, is initially zero. Later, as a result of the decimation, the constraints imposed by the channel and energy minimization, it will be assigned a value.

\begin{figure}[htbp]
\begin{center}
\leavevmode
\includegraphics[width=60mm,keepaspectratio]{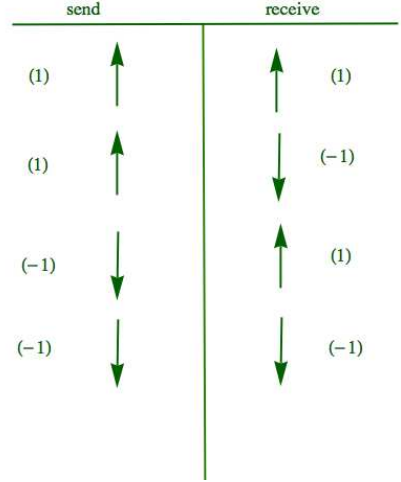}
\caption{The possible states of the basic unit (one send, one receive) and corresponding numerical values.}
\label{sendrcv}
\end{center}
\end{figure}

The four possible states of a basic unit are listed in figure \ref{sendrcv}.
\noindent For a concrete example, $H_{\uparrow \downarrow}$ is computed as
$$
-J(1\ast -1)+ h(1+ -1)= J.
$$
\noindent The remaining energies are computed similarly. 

The following conventions and definitions are standard and will be referred to in the sequel. Let $\beta$, defined as the inverse of the send/receive unit system temperature, be constant (isothermal). The Helmholtz free energy of the system, ${\cal F}$, is defined by
$$Z=e^{-\beta {\cal F}}$$
\noindent where the partition function (the sum over all configurations) is given by
$$
Z=e^{-\beta  H_{\uparrow  \uparrow} }+ e^{-\beta  H_{\uparrow  \downarrow}}+e^{-\beta  H_{\downarrow  \uparrow} }+ e^{-\beta  H_{\downarrow  \downarrow}}.
$$

\noindent The probability of any particular spin configuration, for example $\uparrow \uparrow$ (send a ``1'', receive a ``1'')
is represented in Boltzmann Gibbs form as
$$
P_{\uparrow \uparrow}= \frac{e^{-\beta  H_{\uparrow  \uparrow} }}{Z}.
$$

The concept of internal energy, denoted $U$, is a {\it macroscopic} observable and a common 
object of study in thermodynamics. Changes in the internal energy , $\Delta U$, are decomposed into two 
distinct macroscopic types: work $\Delta W$ and heat $\Delta Q$. As noted in the introduction, 
macroscopic descriptions of work are varied and system dependent. Thermodynamic 
expressions of Gibbs free energy often inherit this system dependent language.
Statistical mechanics offers an additional point of view. In terms of the {\it microscopic} 
data, the internal energy  associated with a basic send/receive unit is also well represented by \cite{hill}, \cite{huang}
$$
U=P_{\uparrow \uparrow} H_{\uparrow \uparrow}+P_{\uparrow \downarrow} H_{\uparrow \downarrow}+
P_{\downarrow \uparrow} H_{\downarrow \uparrow}+P_{\downarrow \downarrow} H_{\downarrow \downarrow}.
$$
\noindent Similarly, the generic form of work in terms of given changes in the microscopic data is  \cite{ford}
$$
\begin{array}{cc}
\Delta W= & P_{\uparrow \uparrow} \Delta H_{\uparrow \uparrow}+P_{\uparrow \downarrow} \Delta H_{\uparrow \downarrow}+\\
& P_{\downarrow \uparrow} \Delta H_{\downarrow \uparrow}+P_{\downarrow \downarrow} \Delta H_{\downarrow \downarrow}.
 \end{array} 
$$

\section{A Trivial Decimation}
A channel is represented mathematically as a matrix of conditional probabilities. For example, the probability
that the receiver observes a $\downarrow$ given that a $ \uparrow$ was sent: $P_{\downarrow | \uparrow}$.
The channel for the send/receive unit system introduced in the last section is simply
$$
\left(
\begin{array}{cc}
 P_{\uparrow | \uparrow} &P_{\downarrow | \uparrow} \\ 
P_{\uparrow | \downarrow} & P_{\downarrow | \downarrow}  
\end{array} 
\right)
$$
\noindent where the rows of the matrix sum to one.
In terms of the Boltzmann Gibbs distribution, the probability of receiving a spin up given that a spin up was sent is

$$P_{\uparrow | \uparrow}=\frac{ e^{-\beta  H_{\uparrow  \uparrow}}}{e^{-\beta  H_{\uparrow  \uparrow} }+ e^{-\beta  H_{\uparrow  \downarrow}}}.$$
The probability of receiving a spin down given that a spin up was sent is
$$P_{\downarrow | \uparrow}=\frac{ e^{-\beta  H_{\uparrow  \downarrow}}}{e^{-\beta  H_{\uparrow  \uparrow} }+ e^{-\beta  H_{\uparrow  \downarrow}}}.$$

\noindent Define the standard \cite{kadanoff} rescaled Hamiltonian, $\tilde{H},$ by summing out (decimation) the spin values at site 2 (the ``receiver'')

$$e^{-\beta  \tilde{H}_{\uparrow}} = e^{-\beta  H_{\uparrow  \uparrow} }+ e^{-\beta  H_{\uparrow  \downarrow}}$$

\noindent so that by the definition of the channel 

$$ \tilde{H}_{\uparrow} = \frac{1}{\beta} Log(P_{\uparrow | \uparrow} ) + H_{\uparrow  \uparrow} $$
\noindent or equivalently

$$ \tilde{H}_{\uparrow} = \frac{1}{\beta} Log(P_{\downarrow | \uparrow} ) + H_{\uparrow  \downarrow}.$$

\noindent In the parlance of communications theory, the ``sender's hamiltonian'', $\tilde{H}$, determines (via Boltzmann Gibbs) the ``sender's marginal", $\tilde{P}$. Notice that the channel is invariant under a transformation

$$
\begin{array}{cc}
& H_{\uparrow  \uparrow}   \rightarrow  H_{\uparrow  \uparrow} + c_{\uparrow} \\ 
& \tilde{H}_{ \uparrow} \;  \rightarrow  \tilde{H}_{\uparrow} \; + c_{\uparrow}.
 \end{array} 
$$

\noindent In a similar way, $\tilde{H}_{\downarrow}$ and $c_{\downarrow}$ are obtained in terms of the original unit system and channel via decimation. 

A single constant is sufficient to arbitrarily adjust the sender's marginal at fixed channel. Without loss of generality set 
$$c_{\downarrow}=-c_{\uparrow}.$$ Interestingly, from a pedagogical point of view, this is the only choice consistent with the assumption of a canonically distributed unit system \cite{ford2}. The renormalized (sender's) partition function, $\tilde{Z}$, is defined
$$
\tilde{Z}=e^{-\beta  \tilde{H}_{\uparrow}}+e^{-\beta  \tilde{H}_{\downarrow}}.
$$

\subsection{Some Quantities of Physical Interest}
The capacity is defined as the (maximized) difference of two information theoretic quantities: the entropy associated with the 
receiver's marginal distribution, $S(\hat{P})$
and the average of the channel entropies according to the sender, $<S_{channel}>_{\tilde{P}}$.
But what is the connection to the physics?

Consider first $<S_{channel}>_{\tilde{P}}$.
This is simply ($-\beta$ times) the work done on the four state system by the $\tilde{H}$ decimation. For example
$$
\tilde{P}_{\uparrow} \left( -P_{\uparrow | \uparrow} Log(P_{\uparrow | \uparrow}) -
P_{\downarrow | \uparrow} Log(P_{\downarrow | \uparrow})\right)
$$
is easily seen to be 
$$
-\beta (P_{\uparrow  \uparrow} (\tilde{H}_{\uparrow} - H_{\uparrow \uparrow}) + P_{\uparrow  \downarrow} (\tilde{H}_{\uparrow} - H_{\uparrow \downarrow}))
$$
\noindent with the $\tilde{P}_{\downarrow}$ terms making a similar contribution.

Secondly, the receiver's entropy, $S(\hat{P})$ is calculated in terms of the sender's marginal $\tilde{P}$ and the channel columns as
$$-
\left[\hspace{-1.5mm} \begin{array}{c}
\scriptstyle{P_{\uparrow | \uparrow} }\\
\scriptstyle{ P_{\uparrow | \downarrow} }\\ 
\end{array} \hspace{-1.5mm} \right]
 \hspace{-0.5mm} \cdot
\scriptstyle{\tilde{P} }\;
\textstyle{\textrm {Log}}
\hspace{-0.5mm}\left(\left[\hspace{-1.5mm} \begin{array}{c}
\scriptstyle{P_{\uparrow | \uparrow} }\\
\scriptstyle{ P_{\uparrow | \downarrow} }\\ 
\end{array} \hspace{-1.5mm} \right]
 \hspace{-1.5mm} \cdot
\scriptstyle{\tilde{P} } \hspace{-0.5mm} \right)\,- \,
\left[\hspace{-1.5mm} \begin{array}{c}
\scriptstyle{P_{\downarrow | \uparrow} }\\
\scriptstyle{ P_{\downarrow | \downarrow} }\\ 
\end{array} \hspace{-1.5mm} \right]
 \hspace{-0.5mm} \cdot
\scriptstyle{\tilde{P} }\;
\textstyle{\textrm {Log}}
\hspace{-0.5mm}\left(\left[\hspace{-1.5mm} \begin{array}{c}
\scriptstyle{P_{\downarrow | \uparrow} }\\
\scriptstyle{ P_{\downarrow | \downarrow} }\\ 
\end{array} \hspace{-1.5mm} \right]
 \hspace{-1.5mm} \cdot
\scriptstyle{\tilde{P} }\hspace{-0.5mm} \right).
$$
Exercise the right to decimate the system over the spin at site 1 and define $\hat{H}$
$$e^{-\beta  \hat{H}_{\uparrow}} = e^{-\beta  H_{\uparrow  \uparrow} }+ e^{-\beta  H_{\downarrow  \uparrow}}$$
$$e^{-\beta  \hat{H}_{\downarrow}} = e^{-\beta  H_{\uparrow  \downarrow} }+ e^{-\beta  H_{\downarrow  \downarrow}}.$$

\noindent The entropy of the receiver's marginal may then be expressed
$$
\beta (P_{\uparrow  \uparrow} \hat{H}_{\uparrow}  + P_{\downarrow  \uparrow} \hat{H}_{\uparrow} 
+P_{\uparrow  \downarrow} \hat{H}_{\downarrow}  + P_{\downarrow  \downarrow} \hat{H}_{\downarrow} +\frac{1}{\beta}Log Z).
$$

\noindent Putting the pieces together, it is seen that the channel capacity, i.e.  the (maximized) quantity
$$S(\hat{P}) \, - <S_{channel}>_{\tilde{P}}$$ is given by the maximum over $c_{\uparrow}$
of
\begin{equation}\label{est}
\beta \sum_{spins} P_{\bullet \bullet}([\tilde{H}_{\bullet}+\hat{H}_{\bullet}]-H_{\bullet \bullet}) +Log Z.
\end{equation}
This is easily seen to be ($-\beta$ times) the change in the Gibbs free energy 
$$\Delta \mathcal{G}=  -\Delta W  + \Delta \mathcal{F}$$ 
\noindent where the change in work, $-\Delta W$  (for a famous, {\it macroscopic} example, $- \left( -p dV \right) \,$) and the change in Helmholtz free energy, $\Delta \mathcal{F}$, are associated with separating a composite {\it microscopic} spin configuration, for example $\uparrow \downarrow$,
into its component spins, $\uparrow$ and $\downarrow$, via decimations. 
By inspection, the work term, $\Delta$W, in equation (\ref{est}) is apparent in its {\it microscopic},
statistical mechanical form ``$p \Delta H$''. A moments reflection identifies the free energy changes. For by the definition of the decimated partition functions $\tilde{Z}$ and $\hat{Z}$, the change
in the Helmholtz free energy, $\Delta \mathcal{F}$, is given by

$$
-\frac{1}{\beta} ([ Log\tilde{Z}+ Log\hat{Z}] -Log Z) = -\frac{1}{\beta} Log Z.
$$

\noindent In this way it is seen that determining the capacity of the two site spin system is equivalent to the standard Gibbs free energy analysis.
The capacity is the maximum amount of work (other than the work done by the sender/receiver decimations) that may be extracted from
the system consistent with the channel constraints.

\subsection{Symmetric Channel}
The prototypical example of a simple communications system is the
binary symmetric channel. In the context of the Ising prototype discussed in this letter, the channel symmetry is
achieved by setting the applied field $h$ equal to zero. The hamiltonian 
$$ H_{\sigma_1 \sigma_2}=-J \sigma_{1} \sigma_{2}$$
depends on the coupling 
term $J$ and
$$
\left(
\begin{array}{cc}
 P_{\uparrow | \uparrow} &P_{\downarrow | \uparrow} \\ 
P_{\uparrow | \downarrow} & P_{\downarrow | \downarrow}  
\end{array} 
\right)
=
\left(
\begin{array}{cc}
 1- \delta & \delta \\ 
\delta & 1 - \delta 
\end{array} 
\right)
$$ 
\noindent where
$$
\delta=\frac{ e^{-\beta  J} }{ e^{-\beta  J}+ e^{\beta  J} }.
$$
As described in the previous section, decimating over the ``receivers spin'' defines the $\tilde{H}$ hamiltonian
$$
\begin{array}{cc}
&\tilde{H}_{\uparrow} = -\frac{1}{\beta} Log\left(  e^{-\beta  J}+ e^{\beta  J}\right) + c_{\uparrow} \\ 
& \tilde{H}_{\downarrow} = -\frac{1}{\beta} Log\left(  e^{-\beta  J}+ e^{\beta  J}\right) - c_{\uparrow}.
 \end{array} 
$$
\noindent Recall that the inclusion of $c_{\uparrow}$ leaves the channel invariant and is sufficient to arbitrarily
adjust  $\tilde{P}$, the senders marginal distribution.

\begin{figure}[htbp]
\begin{center}
\leavevmode
\includegraphics[width=50mm,keepaspectratio]{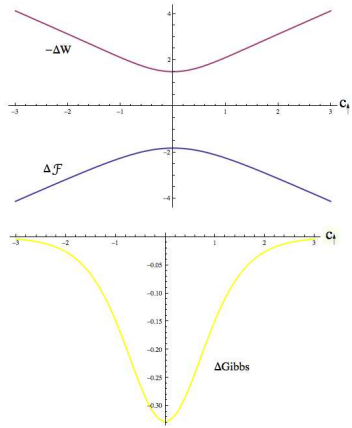}
\caption{Upper frame: the work and free energy changes associated with separating the composite spin states into 
components at $\beta=1$, $J=1$. Lower frame: the changes in Gibbs free energy as $c_{\uparrow}$ is varied, i.e. extremization over the sender's marginal distribution.}
\label{deltagibbs}
\end{center}
\end{figure}

Obtain $\hat{H}$  similarly, i.e. by decimation over the ``senders spin'' and respecting the channel constraints
$$
\begin{array}{cc}
& \hat{H}_{\uparrow} = -\frac{1}{\beta} Log\left(  e^{-\beta  (J - c_{\uparrow})}+ e^{\beta ( J- c_{\uparrow})}\right) \\ 
& \; \hat{H}_{\downarrow} = -\frac{1}{\beta} Log\left(  e^{-\beta ( J+c_{\uparrow})}+ e^{\beta  (J+c_{\uparrow})}\right).
\end{array} 
$$
With these quantities in hand, the Gibbs free energy changes consistent with the channel constraints
may be computed, following equation (\ref{est}), as a function of $c_{\uparrow}$. Figure \ref{deltagibbs} shows the results for the case
$\beta=1$, $J=1$. With these parameter settings and accompanying channel constraints, the minimum
$\Delta G$, i.e. $-1$ times the channel capacity, is achieved at $c_{\uparrow}=0$.
\section{summary}
The binary channel is simultaneously a basic example of a communications system often studied in information 
theory and a basic example of a magnetic spin system. A straightforward comparison of the associated channel capacity and Gibbs
free energy procedures finds no difference (other than multiplication by -$\beta$)  between the two concepts.


\section{Bibliography}

\begin{thebibliography}{10}
\bibitem{gibbs}
J.W. Gibbs,
\newblock  ``On the Equilibrium of Heterogeneous Substances",
\newblock{Transactions of the Connecticut Academy of Arts and Sciences 3},
1876

\bibitem{callen}
H. Callen,
\newblock  ``Thermodynamics and an Introduction to Thermostatistics",
\newblock{Wiley},
1985

\bibitem{l&l}
L. Landau; E. Lifshitz,
\newblock  ``Theory of Elasticity",
\newblock{Pergamon, Oxford},
1986

\bibitem{krevet}
B. Krevet, M. Kohl, P. Morrison, S. Seelecke,
\newblock  ``Magnetization and strain dependent free energy model for FEM simulation of magnetic shape memory alloys",
\newblock{Eur. Phys. J. Special Topics, 158},
2008

\bibitem{kim}
S.J. Kim,
\newblock  ``A rate dependent thermo-electro-mechanical free energy model for perovskite type single crystals",
\newblock{International Journal of Engineering Science, 45},
2007

\bibitem{s&w}
C. Shannon; W. Weaver,
\newblock  ``The Mathematical Theory of Communication",
\newblock{University of Illinois Press},
1949

\bibitem{jaynes}
E.T. Jaynes,
\newblock  ``Information Theory and Statistical Mechanics",
\newblock{Phys. Rev., 106, 620},
1957

\bibitem{feynman}
R. Feynman, M. Sands,
\newblock  ``The Feynman Lectures on Physics, vol. 3",
\newblock{Addison Wesley},
1998


\bibitem{b&k}
T. Burkhardt; W. Kinzel,
\newblock  ``Real-space renormalization for structural phase transitions in the Ising universality class",
\newblock{Physical Review B, Volume 20, Issue 11},
1979

\bibitem{chin&hong}
Chin-Kun Hu, Hong-Yuh L
\newblock  ``Renormalization-Group Study of a Simple Cubic Ising Model with Four-Spin Interaction",
\newblock{Chinese Journal of Physics, Volume 22, No. 3},
1984

\bibitem{f&h}
J. Fortin, P. Holdsworth,
\newblock  `` Real space renormalization group analysis of the random field Ising model",
\newblock{J. Phys. A: Math. Gen. 29 L539-L545},
1996

\bibitem{nishimori}
H. Nishimori,
\newblock  `` Optimum Decoding Temperature for Error-Correcting Codes",
\newblock{J. Phys. Soc. Jpn. 62  pp. 2973-2975},
1993

\bibitem{n&w}
H. Nishimori, K. Wong,
\newblock  `` Statistical mechanics of image restoration and error-correcting codes",
\newblock{Phys. Rev. E 60, 132 - 144},
1999

\bibitem{tanaka}
T. Tanaka,
\newblock  ``Statistical mechanics of CDMA multiuser demodulation",
\newblock{Euro- 
phys. Lett., vol. 54, pp. 540Ð546},
2001

\bibitem{kadanoff}
L. Kadanoff,
\newblock  `` Statistical Physics, Statics, Dynamics and Renormalization",
\newblock{World Scientific},
2000


\bibitem{hill}
T.L. Hill,
\newblock  `` Statistical Mechanics",
\newblock{McGraw- Hill Book Co., New York},
1956

\bibitem{huang}
K. Huang,
\newblock  ``Statistical Mechanics",
\newblock{Wiley},
1987

\bibitem{ford}
D. Ford,
\newblock ``Surfaces of Constant Temperature in Time,''
\newblock {http://www.arxiv.org/abs/cond-mat/0510291},
2005

\bibitem{ford2}
D. Ford,
\newblock ``Application of Thermodynamics to the Reduction of Data Generated by a Non-Standard System,''
\newblock {http://arxiv.org/abs/cond-mat/0402325v1},
2004



\end{thebibliography}
\end{document}